\documentclass[twocolumn,showpacs,preprintnumbers,amsmath,amssymb]{revtex4}

\usepackage{graphicx}
\usepackage{dcolumn}
\usepackage{bm}
\usepackage{here}

\begin{document}

\title{Quantum dynamics and delocalization in coherently driven
one-dimensional double-well system
}

\author{Akira Igarashi}
 \email{f99j806b@mail.cc.niigata-u.ac.jp}
 \affiliation{
 Graduate School of Science and Technology, 
 Niigata University, Ikarashi 2-Nochou 8050, Niigata 950-2181, Japan
 }
 
 \author{Hiroaki S. Yamada}
 \email{hyamada@uranus.dti.ne.jp}
 \affiliation{%
 YPRL, 5-7-14 Aoyama, Niigata 950-2002, Japan
 }%
 \date{\today}

  \begin{abstract}
  We numerically study influence of a polychromatic perturbation 
  on wave packet dynamics 
  in one-dimensional double-well potential.
  It is found that time-dependence of the transition probability between the wells shows
  two kinds of the motion typically,  coherent oscillation and irregular fluctuation 
combined to the delocalization of the wave packet, 
  depending on the perturbation parameters.
  The coherent motion changes the irregular one 
  as the strength and/or the number of frequency components of
  the perturbation increases. We  discuss a relation between our model and decoherence
  in comparing with the result under stochastic perturbation.
  Furthermore we compare the quantum fluctuation,
  tunneling in the quantum dynamics with ones in the semiquantal dynamics.
 \end{abstract}

   
   \maketitle

\section{Introduction}
Quantum fluctuation can suppress chaotic motion of wave packet in the phase space
due to the quantum interference, as seen in kicked rotor \cite{gutzwiller90,hu98}. 
On the contrary, the quantum fluctuation can enhance the chaotic motion 
of wave packet due to tunneling effect
as seen in kicked double-well model \cite{pattanayak94,zhang95}.
The relation between chaotic behavior and tunneling phenomenon 
in classically chaotic systems 
is interesting and important subject in study of quantum physics \cite{razavy03,tomsovic98}.
Recently, the semiclassical description for the tunneling phenomena in a
classically chaotic system have been developed by
several groups \cite{shudo98,takatsuka99,brodier02}. 

Lin and Ballentine studied interplay between the tunneling and classical chaos for a particle in a
double-well potential with oscillatory driving force \cite{lin90}. 
They found that coherent tunneling
takes place between small isolated classical stable regions of phase space bounded by 
Kolmogorov-Arnold-Moser (KAM) surfaces, which are much smaller than the volume of a 
single potential well.  

H\"{a}nggi and the coworkers studied the chaos-suppressed tunneling in the driven double-well model 
in terms of the Floquet formalism \cite{grossmann91}. 
   They found a one-dimensional manifold in the parameter space, where the tunneling completely
   suppressed by the coherent driving.
   The time-scale for the tunneling between the wells diverges 
   because of intersection of the ground state doublet  of the quasienergies. 
   
   While the mutual influence of quantum coherence and classical chaos has been under investigation since
   many years ago, the additional effects caused by coupling the chaotic system to 
the other degrees of freedom (DOF)
   or  an environment, 
   namely {\it decoherence and dissipation}, have been studied only rarely \cite{grossmann91,weiss99} as well as 
   the tunneling phenomena in the chaotic system.
   Since mid-eighties there are some studies on environment-induced quantum
   decoherence by coupling the quantum system to a reservoir
   \cite{leggett81,dittrich86,choen99}.
   Recently quantum dissipation due to the interaction with chaotic DOF 
has been also studied\cite{zurek81,choen97,kolovsky94}.

   In this paper we numerically investigate the relation {\it quantum fluctuation, tunneling and decoherence}
combined to the delocalization  in 
   wave packet dynamics in  one-dimensional double-well system driven by 
   polychromatic external field.

Before closing this section, we refer to a study on a delocalization 
phenomenon by a perturbation with some frequency components in the other model.
Casati {\it et al.} have reported that the kicked rotator model with a frequency
 modulation amplitude of kick can be mapped to the tight-binding 
form (Loyld model) on higher-dimensional lattice in solid-state physics 
under very specific condition \cite{casati89,borgonovi97}. Then the number $M$ of the incommensurate 
frequencies corresponds the dimensionality of the tight-binding system. 
The problem can be efficiently reduced to a localization problem in $M+1$ 
dimension. As seen in the case of kicked rotators, we can also expect that 
in the double-well system the coupling with oscillatory perturbation 
is roughly equivalent to an increase in effective degrees of freedom and 
a transition from a localized wave packet to delocalized one is enhanced 
by the polychromatic perturbation. The concrete confirmation of the naive 
expectation is one of aims of this numerical work.

We present the model in the next section. 
In Sect.3, we show the details of the numerical results of the time-dependence of the 
transition probability between the wells based on the quantum dynamics. 
Section 4 contains the summary and discussion. 
Furthermore, in appendix A, we gave details of the classical phase space portraits in the 
polychromatically perturbed double-well system and some considerations to the effect of 
polychromatic perturbation.
In appendix B, a simple explanation 
for the perturbed instanton tunneling picture
is given.

   
\section{Model}
  We consider a system described by the following Hamiltonian, 
   \begin{eqnarray}
    H (t) & = & \frac{p^2}{2} + \frac{q^4}{4} -A(t) \frac{q^2}{2},  \\
    A(t) & =& a-\frac{1}{\surd M} \sum_{i=1}^M \epsilon_i \sin(\Omega_{i}  t +\theta_i ).   
   \end{eqnarray}
   For the sake of simplicity, $\{ \epsilon_i \}$ and $\{ \theta_i  \}$ are 
   taken as $\epsilon_i=\epsilon$, $\theta_i=0$, $i=1,2,.., M$ in the present paper.  
   Then $M$ is the number of frequency components of the external field and 
   $\epsilon$ is the perturbation strength respectively. 
   \{$\Omega_{i}$\} are order of unity and mutually incommensurate
   frequencies.  We choose off-resonant frequencies which are 
   far from both classical and quantum resonance
   in the corresponding unperturbed problem.
   The parameter $a$ adjusts the distance between the wells
   and we set $a=5$ to make some energy doublets below the potential barrier.
   Note that Lin {\it et al.} dealt with a double-well system driven by forced oscillator
(Duffing-like model), 
   therefore, the asymmetry of the potential plays an important  role in the
   chaotic behavior and tunneling transition between 
   the symmetry-related KAM tori \cite{lin90,grossmann91}.
   However, in our model the potential is remained symmetric during the time evolution process,  
   and different mechanism from the forced oscillation makes 
   the classical chaotic behavior \cite{yamaguchi85,holmes83,duffing}. 
   
   In the previous paper \cite{igarashi05} we presented 
   numerical results concerning 
   a classical and quantum description of the field-induced barrier
   tunneling under the monochromatic perturbation ($M=1$). 
   In the unperturbed double-well system ($M=0$) the instanton describes 
   the coherent tunneling motion of the initially localized wave packet.
   It is also shown that the monochromatic perturbation can breaks the coherent motion 
   as the perturbation strength increases near the resonant frequency
 in the previous paper.
In the classical dynamics of our model, outstanding feature different from previous studies is  
parametric instability caused by the polychromatic perturbation. 
   
   Based on our criterion given below, 
   we roughly estimate the type of the motion, i.e. the coherent and irregular motions,   
   in a regime of the parameter space spanned by the amplitude and the
   number of frequency components 
   of the oscillatory driving force.
 It is suggested that the occurrence of the irregular motion is related to 
dissipative property which is organized in the quantum physics 
   \cite{yamada99}.
The classical phase space portraits and  simple explanation of relation 
to the  dissipative property are given in appendix A.

\section{Numerical results} 
We use Gaussian wavepacket with zero momentum as the initial state, 
which is localized in the right well of the potential. 
\begin{eqnarray}
  \psi (q, t=0) =  (\sigma \pi)^{1/4} \exp \{ -\frac{(q-q_0)^2}{2\sigma}  \}, 
\end{eqnarray}
where $q_0=\surd a \simeq 2.236 $ means a bottom of the right well.  
 The Gaussian wavepacket can be approximately generated by the linear combination
of the ground state doublet as 
$ \psi (q, t=0) \simeq \frac{1}{\surd 2}(\varphi_0(q) + \varphi_1(q)  )$, where
$\varphi_0$ and $\varphi_1$ denote the ground state doublet. 
The recurrence time for the wavepacket is 
$T \equiv 2\pi \hbar/ \Delta \epsilon_{01} \sim 9.4 \times 10^3$ in the unperturbed case ($M=0$), 
where $\Delta \epsilon_{10}$ is the energy difference between the tunneling doublet 
of the ground state. 
We set the spread of the initial packet $\sigma=1/3.4 (\sim 0.3)$ and  $\hbar=1.0$ for simplicity 
throughout this paper. 
Indeed, the ammonia molecule is well described by 
two doublets below  the barrier heigth in unperturbed case.

   We numerically calculate the solution $\psi(q,t)$ 
   of time-dependent Schr\"{o}dinger equation by using second order unitary integration with
   time step $\delta \sim 10^{-2}$.
   We define {\it transition probability} of finding the wave packet in the left well,  
   \begin{eqnarray}
    P_{L}(t) \equiv \int_{-\infty}^{0}|\psi(q,t)|^{2}\,dq. 
   \end{eqnarray}
   In the cases that the perturbation strength is relatively small, 
   $P_L(t)$ can be interpreted as the tunneling probability that the initially
   localized wave packet goes through the central energy barrier and reaches
   the left well.
   We can expect that the transition probability $P_L(t)$ is enhanced as 
   the number $M$ of the frequency components increases up to some extent 
   because of the increasing of the stochasticity in the total system. 
  
  \begin{figure}[!ht]
   \centering
  \includegraphics[clip,scale=1.2]{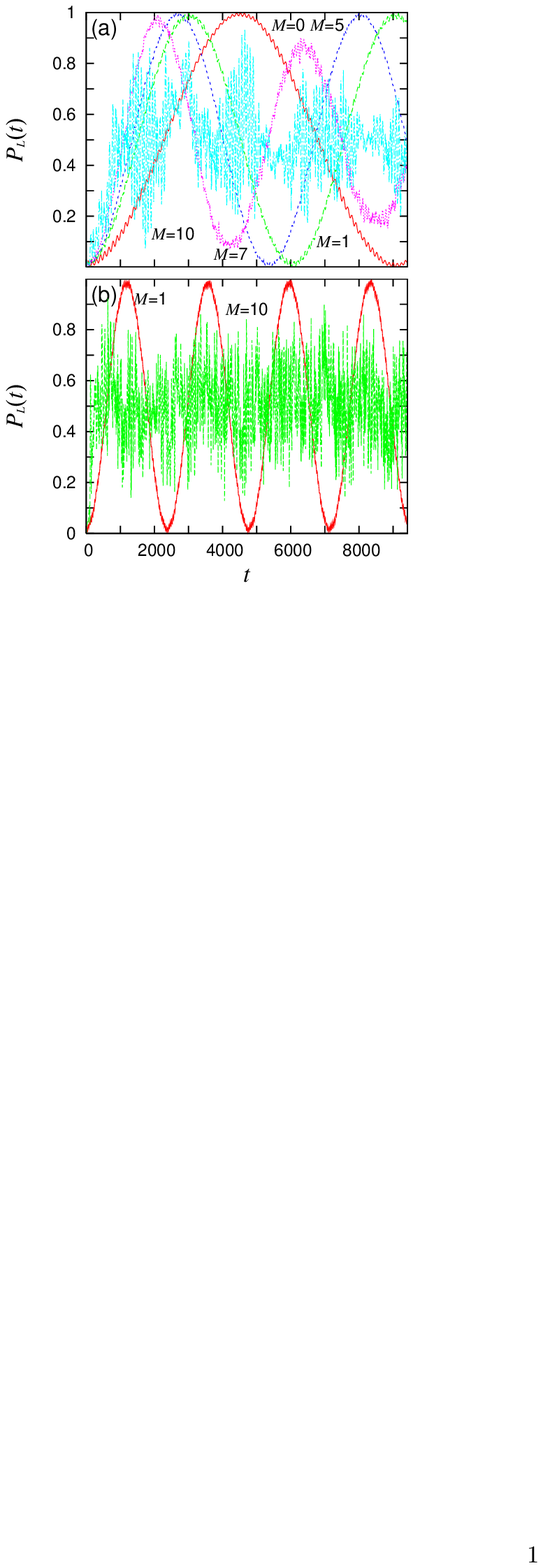}
    \caption{(Color online) Tunneling probability $P_{L}(t)$ as a function of time $t$
    for various $M$'s. (a)$\epsilon=0.4$. (b)$\epsilon=0.8$.
    The calculation time is same order to Heisenberg time in the
    unperturbed case.
    }
   \end{figure}

   Figure 1 shows the time-dependence of 
   $P_{L}(t)$ for various combinations of $\epsilon$ and $M$.
   Apparently we can observe the coherent and irregular motions.
   The coherent motion of the wave packet
   can be well-described by the semiquantal picture in a sense that the wave packet
does not delocalize to the fully delocalized state.
   The semiquantal  picture decomposes the motion of the wave packet into 
   {\it evolution of the centroid motion} and {\it the spreading and squeezing} 
   of the packet \cite{pattanayak94}. (See Subsect.3.5.)
   For example, in cases of relatively small perturbation strength ($\epsilon=0.4$),
   coherent motion remains still  up to relatively large $M(=5)$. 

   It is important to emphasize that the tunneling contribution to the transition 
probability $P_L$ is not so significant for large $\epsilon$ and/or $M$. 
Then $P_L$ may be interpreted as a barrier crossing probability due to the activation-transition 
 because the energy of wave packet increases over the barrier height in the parameter range. 
Especially, in the relatively large perturbation regime
we can interpret the delocalized states as 
chaos-induced delocalization in a sense that the classical 
chaos enhances the quantum barrier crossing rate quite significantly.
The chaotic behavior in the classical dynamics is 
given in appendix A, based on the classical Poincar\'e
section and so on \cite{igarashi06}. In the present section, we mainly focus on the transition
of the quantum state from the localized wavepacket to delocalized state based on the data of   
numerical calculation.

 \subsection{decoherence of the dynamics}  
   Once the wave packet incoherently spreads into the space 
   as the $M$ and/or $\epsilon$ increase, 
the wavepacket is delocalized and never return to Gaussian shape again
within the numerically accessible time.
   Apparently, we regard the delocalized quantum state as a decoherent state
   in a sense that the behavior of the wave packet is similar to that of 
   the stochastically perturbed case. (See Fig.5(a).) 
   In case of relatively small 
   perturbation strength ($\epsilon=0.4$), the decoherence of quantum dynamics appears 
   at around $M \simeq 7$, and  $P_L(t)$ fluctuates irregularly in case of large $M(=10)$.
In short, the irreversible delocalization of a Gaussian wave packet generates 
a transition from coherent oscillation to irregular fluctuation of $P_L(t)$.
   We have confirmed that the similar
   behavior is also observed for other sets of values of the frequencies
   and the different initial phases $\{ \theta_i \}$ of the perturbation.
   
  Here, we define a {\it degree of coherence} 
   $\Delta P_{L}$ of the time-dependence of $P_L(t)$, 
based on the fluctuation of the  transition probability
   in order to estimate quantitatively the difference 
   between coherent and incoherent motions.
   \begin{eqnarray}
    \Delta P_{L} \equiv 
     \surd \langle (P_{L}(t)-\langle P_L(t) \rangle_{T})^{2} \rangle _T, 
   \end{eqnarray}
   where $\langle ... \rangle _{T}$ represents time average value for a period 
   $T=9.4 \times 10^{3}$.
  Note that we used $\Delta P_{L}$ in order to express the decoherence of the
    tunneling osccilation of the transition probability 
in the parametrically perturbed double-well system. 
On the other hand,
    the other quantities such as {\it purity}, {\it linear entropy} and
    {\it fidelity}, are sometimes used to characterize the decoherence
    of the quantum system \cite{prosen02}.
The transition of the dynamical behavior based on
    the fidelity for description of the
    decoherence in the double-well system will be given elsewhere
    \cite{igarashi06}.

   \begin{figure}[!ht]
    \centering
    \includegraphics[clip,scale=1.2]{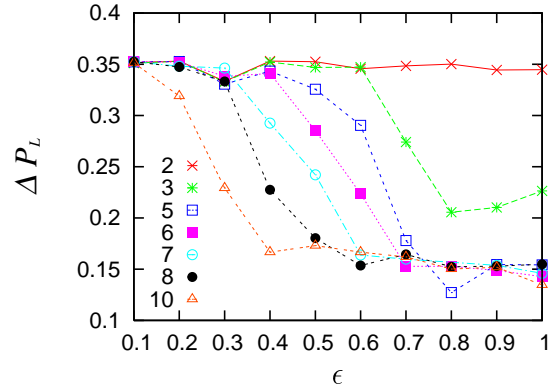}
    \caption{Perturbation strength $\epsilon$ dependence of the degree of coherence  $\Delta P_{L}$ of 
    tunneling probability for various $M$'s.
    The $\Delta P_L$ is numerically estimated by $P_L(t)$.}
   \end{figure}

   Figure 2 shows the perturbation strength dependence of 
   $\Delta P_{L}$ for various $M$'s. 
   We roughly divide the type of  motion of wave packet into three ones as follows.
   In the {\it coherent motions}, the value of $\Delta P_{L}$'s is
   almost same to the unperturbed case, i.e. $\Delta P_{L} \gtrsim 0.3$,  
  in which cases the instanton-like picture is valid \cite{instanton}.
A simple explanation of  the perturbed instanton is given in appendix B.

   In the {\it irregular motions} which are similar to the stochastically perturbed case, 
   the value of $\Delta P_{L}$'s becomes much smaller, i.e. 
   $\Delta P_{L} \lesssim 0.2$. 
   As a matter of course, there are the intermediate cases between the coherent and the irregular motions, 
   $0.2 \lesssim \Delta P_{L} \lesssim 0.3$.
   Note that the exact criterion of the intermediate motion is not important in the present paper
because we can expect that the transitional cases approach to the irregular case in the long-time behavior.
It should be stressed that the critical value $\epsilon_c$ exists, which divides the behavior of $P_L(t)$ 
into regular and irregular motions. 

   \begin{figure}[!ht]
    \centering
    \includegraphics[clip,scale=1.2]{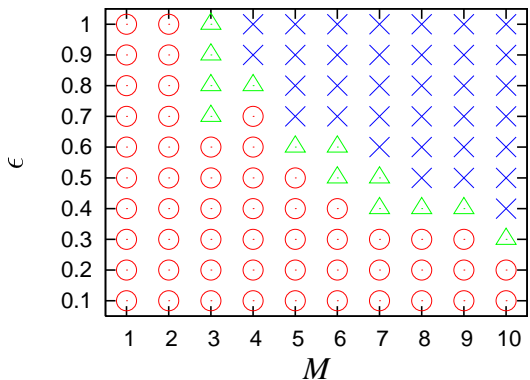}
    \caption{(Color online)
    Approximate phase diagram 
    in the parameter space spanned by the amplitude and the number of frequency components
    of the oscillatory driving force. 
    The motions are plotted as different marks based on the degree of coherence $\Delta P_L$.
    Circles($\bigcirc$), crosses($\times$), and triangles($\triangle$)
    denote coherent motions ($\Delta P_L \geq 0.3$), 
    irregular motions ($\Delta P_L \leq 0.2$), and 
    the transitional cases ($ 0.2 \leq \Delta P_L \leq 0.3$)  respectively.}
   \end{figure}
 
  Figure 3 shows a classification of the motion in the parameter space
   which is estimated by the value of the degree of coherence $\Delta
   P_{L}$. 
   It seems that two kinds of the motion, i.e. coherent and irregular motions, 
   are divided by the thin layer corresponding to the "transitional case".
   As $M$ increases, decoherence of the motion appears 
   even for small $\epsilon$.
   The numerical estimation suggests that there are 
   the critical values $\epsilon_c(M)$ of the 
   perturbation strength depending on $M$. When the perturbation strength
   $\epsilon$ exceeds the critical value $\epsilon_{c}(M)$ for some $M$,
   the tunneling oscillation loses the coherence.  
The approximated phase diagram roughly same as the diagram generated by maximal
Lyapunov exponent of the classical dynamics. 
(See appendix A.)

\subsection{reduction of the tunneling period for $\epsilon < \epsilon_c$}
In this subsection we give a consideration to the reduction of the tunneling
period in the regular motion regime $\epsilon < \epsilon_c$. 

   \begin{figure}[!ht]
    \centering
    \includegraphics[clip,scale=1.2]{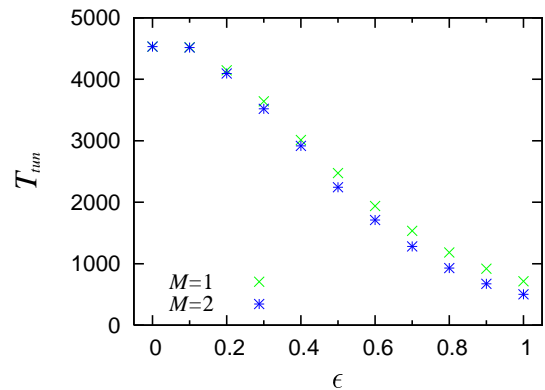}
    \caption{
Perturbation strength dependence of tunneling period $T_{tun}$ for 
$M=1$ and $M=2$ in the coherent motion regime $\epsilon < \epsilon_c$.
    }
   \end{figure}

Figure 4 shows the $\epsilon-$dependence of the 
the period $T_{tun}$ of the tunneling oscillation estimated by the numerical data $P_L(t)$
in the coherent motion regime $\epsilon < \epsilon_c$.
We can observe the monotonically decreasing of the tunneling period 
as the perturbation strength increases.
 In the monochromatically perturbed case,  the reduction of the tunneling period can be 
interpreted by  applying the Floquet theorem to the quasi-energy
states and the quasi-energy as the Hamiltonian is time-periodic $H(t+2\pi/\Omega)=H(t)$.
When the wave packet does not effecively absorb the energy from the external perturbation
the time-dependence of the quantum state can be described by the
linear combination of 
a doublet of quasi-degenerate ground states with opposite parity
 because we prepare the initial state
in $ \psi (q, t=0) \simeq \frac{1}{\surd 2}(\varphi_0(q) + \varphi_1(q)  )$  
and the evolution is adiabatic.
In the two-state approximation that the avoided crossing of 
the eigenvalues dynamics does not appear during the time evolution, 
it is expected that the state evolves as, 
    \begin{eqnarray}
\psi (q, t) \simeq \frac{1}{\surd 2} 
\Bigl( u_0(q,t) e^{-i \int_0^t E_0(\tau) d\tau} + u_1(q,t)e^{-i \int_0^t E_1(\tau) d\tau}  \Bigr),
   \end{eqnarray}
where $(E_0(t), E_1(t))$ and $(u_0(t), u_1(t))$ denote the quasi-energies and Floquet states
of the time-periodic Hamiltonian \cite{holthaus92}. 
Under the approximation we expect the following relation,
    \begin{eqnarray}
T_{tun} \simeq \frac{\pi}{ \Delta E_{01}}, 
   \end{eqnarray}
where $\Delta E_{01}$ means quasi-energy splitting of the ground state doublet
due to the tunneling between the wells.

   \begin{figure}[!ht]
    \centering
    \includegraphics[clip,scale=1.2]{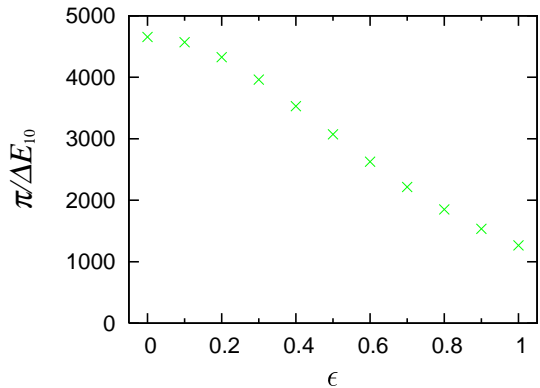}
    \caption{
Perturbation strength dependence of quasi-energy splitting 
 in the coherent motion regime $\epsilon < \epsilon_c$ 
in the monochromatically perturbed case ($M=1$).
    }
   \end{figure}

Let us confirm the relation in Eq.(7) numerically.
In Fig. 5 we show the $\epsilon-$dependence 
of $\pi/\Delta E_{01}$. The behavior is analogus to the 
$\epsilon-$dependence of the tunneling period of the oscillation $P_L(t)$ in Fig. 4, 
 in the weak perturbation regime.

The similar correspondence between the tunneling
period and the change of the
quasi-energy splitting have been reported for the other double-well system 
by Tomsovic {\it et al.} \cite{tomsovic94,kohler98,bonci98,zakrzewski98}. 
It is well-known that the chaos around the separatrix contributes to
 the enhancement of the tunneling split between
the doublet, i.e. chaos-assisted tunneling. 
 The reduction of the tunneling period can be approximately explained by 
the chaos-assisted instanton picture in the coherent oscilation regime $\epsilon <\epsilon_c$. 
The simple explanation for the perturbed instanton picture based on 
the width of the chaotic layer in the classical dynamics is given in 
appendix B. (See also appendix A.)

Generally speaking, as the number of frequencies $M$ increases the tunneling period is 
more reduced as seen in Fig.4 although we do not have analytic representation in the 
polychromatically perturbed cases.
We conjecture that as seen in appendix A 
the increasing of the width of the stochastic layer contributes the reduction of 
the tunneling period even in the polychromatically perturbed cases.

   \begin{figure}[!ht]
    \centering
    \includegraphics[clip,scale=0.7]{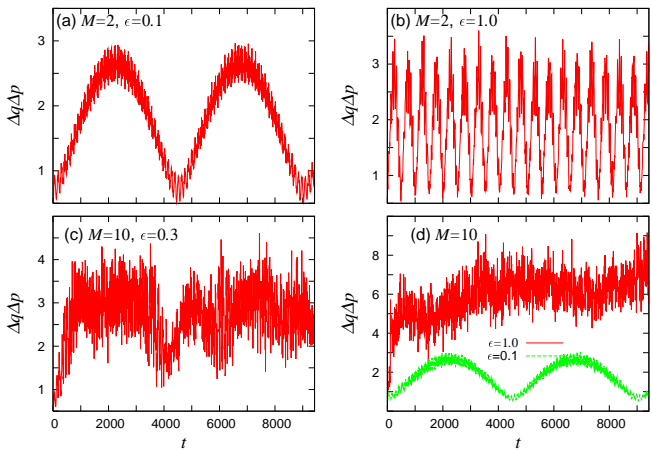}
    \caption{Plots of the uncertainty product $ \Delta q \Delta p $ as a function of time 
    for some combinations of the parameters. 
    (a) $M=2$, $\epsilon=0.1$. (b) $M=2$, $\epsilon=1.0$.
    (c) $M=10$, $\epsilon=0.3$. (d) $M=10$, $\epsilon=1.0$.
    }
   \end{figure} 

   \begin{figure}[!ht]
    \centering
    \includegraphics[clip,scale=0.7]{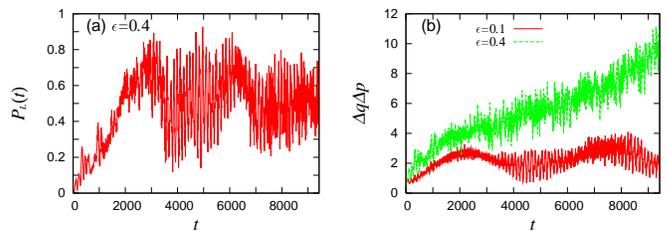}
    \caption{(a)Tunneling probability $P_{L}(t)$ as a function of time $t$
    under stochastic perturbation with $\epsilon=0.4$. 
    (b)Plots of the uncertainty product $ \Delta q \Delta p $ versus time 
    for various $\epsilon'$s with the stochastic perturbation.
    The stochastic perturbation strength $\epsilon$ 
    is normalized to be equivalent to one 
    of the polychromatic perturbation. }
   \end{figure} 
   
 \subsection{uncertainly product}  
 Here let us investigate the spread of the wave packet in the phase space ($q,p$).
  Hitherto we mainly investigated the dynamics in $q-$space by $P_L(t)$.
The phase space volume gives a part of the compensating information for the phase space
dynamics of the wave packet.
 Figure 6 presents the uncertainly product, i.e. phase space volume, 
   as a function of time for various cases, which is
   defined by,  
   \begin{eqnarray}
    \Delta q \Delta p \equiv  \surd \langle (q-\langle q \rangle )^2 \rangle
     \surd \langle (p-\langle p \rangle )^2\rangle,   
   \end{eqnarray}
   where $\langle...\rangle $ denotes quantum mechanical average.
   The uncertainty product can be used as a measure of quantum fluctuation \cite{products}.
   The initial value is $\Delta q \Delta p =\hbar/2(=0.5)$ for the 
   Gaussian wave packet.
   It is found that in the case $M=2$ the increase of the perturbation strength 
   does not break the coherent oscillation and  
   enhances the frequency of the time-dependence of the uncertainty
   product.
   For the relatively large $\epsilon$ in $M=10$, $\Delta q \Delta p$
   increases until the wave packet is relaxed in the space, and it can not return to 
Gaussian wave packet anymore.
   For the larger time scale, it fluctuates around the corresponding certain level. 
   We can expect that the structure of the time dependence  well
   corresponds to the behavior of the transition probability $P_L(t)$ in Fig. 1.

 \subsection{stochastically perturbed case}  
  It will be instructive to compare the above irregular motion 
   under the polychromatic perturbation 
   with the stochastically perturbed one.
   We recall that the stochastic perturbation, composed of the infinite number of the
   frequency components ($M \to \infty$) with absolute continuous
   spectrum, can break the coherent dynamics. 
   Indeed, if the time dependence of the potential 
   comes up with the stochastic fluctuation as 
   $\langle (A(t_1)-a)(A(t_2)-a)\rangle_{en} \propto T_B \delta(t_2-t_1)$, 
   where $\langle ...\rangle_{en}$ and $T_B$ denote ensemble average and
   the temperature respectively, the stochastic perturbation partially models 
   a heat bath coupled with the system  \cite{gammaitoni98}.
   Then the number of the frequency component corresponds to 
   the number of degrees of freedom
   coupled with the double-well system.
   The $P_L(t)$ for the stochastic
   perturbation is shown in Fig. 6(a). 
   The stochastic perturbation can be achieved numerically by replacing $A(t)-a$ in the
   Eq.(2) by random number and we use uniform random number which is
   normalized so that the power of the perturbation is the
   same order to one of the polychromatic case.
   In the limit
   of large $M$ the motion under the polychromatic perturbation tends to
   approach the one driven by the stochastic perturbation
   provided with the same perturbation strengths $\epsilon$.
   
   Figure 7(b) shows the uncertainty product $\Delta q \Delta p$ for the stochastically
   perturbed cases.
   It is found that the time-dependence of the uncertainty product
   in the stochastically perturbed case
   behaves similarly to the polychromatically perturbed ones 
   for the relatively small $\epsilon(\sim 0.1)$.
   On the other hand, for the relatively larger $\epsilon$(=0.4) the time-dependence shows quite different behavior.
   While  $\Delta q \Delta p$ grows linearly with time in the stochastically perturbed case, 
   in the polychromatically perturbed cases the growth of   $\Delta q \Delta p$ 
   saturates at a certain level.
The linear growth of $\Delta q \Delta p$ shows 
that  the external stochasticity breaks the quantum interference in the internal dynamics.
   The growth of $\Delta q \Delta p$ is strongly related to the growth of the
   energy of the packet \cite{igarashi06}.
   In the polychromatically perturbed cases the energy growth saturates at certain level
   due to quantum interference.
   On the other hand, in the case the energy grows unboundedly,  the
   activation transition becomes much
   more dominant than the tunneling transition when the wave packet transfers the opposite well.
   The details concerning relation between the stochastic resonance \cite{gammaitoni98} and 
   suppression of the energy growth will be given elsewhere \cite{igarashi06}. 
   
Note that the polychromatic perturbation can be identified with a white noise 
(or a colored noise if the frequencies are distributed over a finite band width)
only in the limit of $M \to \infty$, while the stochastic perturbation can model a heat bath
that breaks the quantum interference of the system. 
A similar phenomenon by the different property of the perturbation has been
observed as "dynamical localization" and 
the "noise-assisted mixing" of the quantum state in the momentum space 
in the quantum kicked rotor model \cite{adachi88}.

   \begin{figure}[!ht]
    \centering
    \includegraphics[clip,scale=0.7]{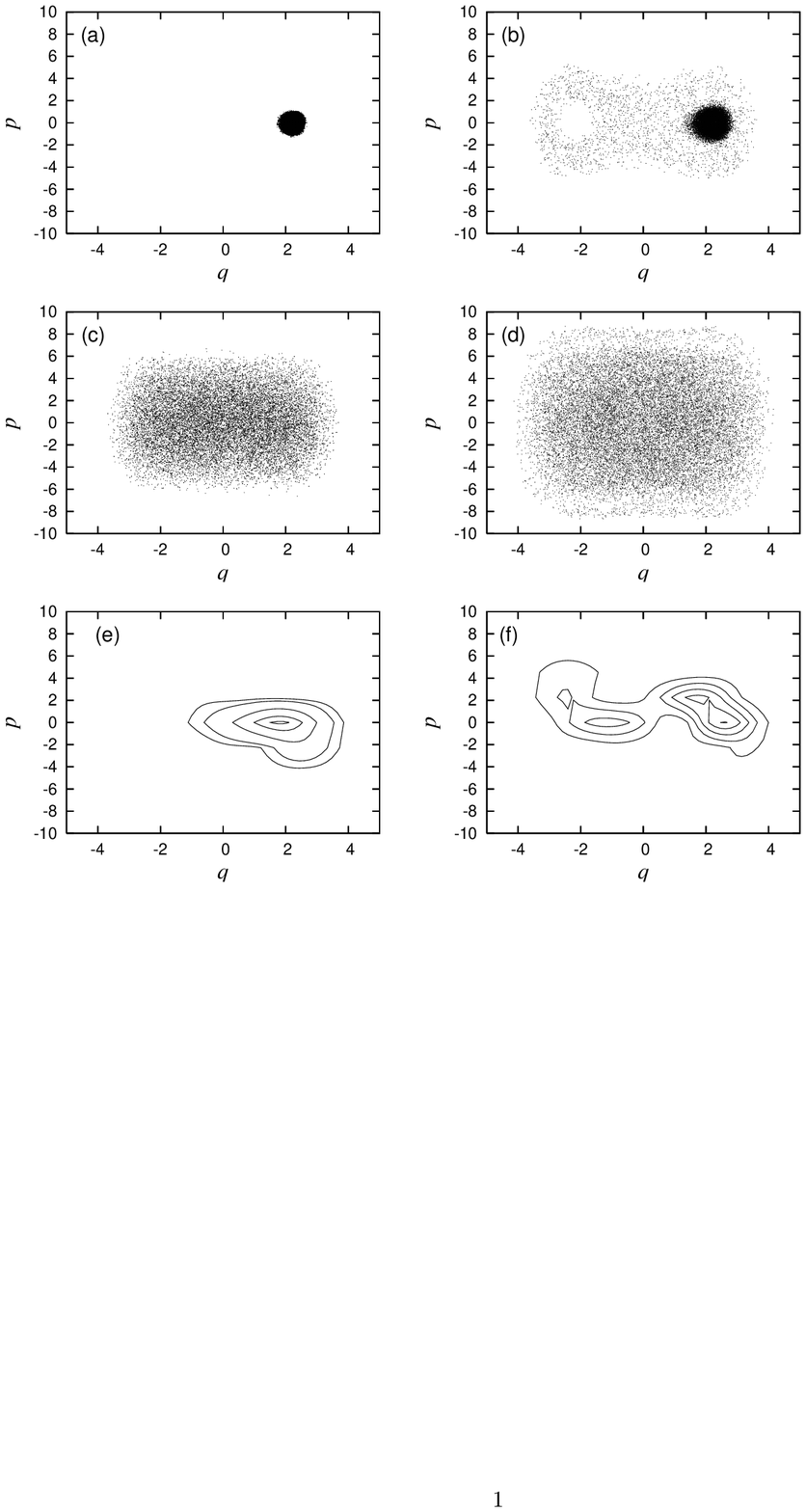}
    \caption{ Classical ((a), (b)) and semiquantal ((c), (d)) 
    stroboscopic phase space portrait 
    of the polychromatically perturbed double-well system
    with the same initial conditions $(q_0,p_0)$ and 
    the same parameters at $t=2\pi n/\Omega_1$, $n=1,2,...$. 
    The selection of initial condition of the 
    fluctuation follows the minimum uncertainty.
    Contour plots ((e), (f)) of the Hushimi functions 
    for the corresponding the quantum state at $t=8.7\times10^3$.
    Contour lines in the panel (e) at the values 0.01, 0.02, 0.05, 0.08 and
    0.1, in the panel (f) at 0.01, 0.02, 0.03, 0.04 and 0.05.
    $M=7$ and  $\epsilon=0.4$ for (a), (c) and (e).
    $M=7$ and  $\epsilon=0.8$ for (b), (d) and (f).
    }
   \end{figure}
 
 \subsection{phase space portrait}  
   Finally, in order to see effect of quantum fluctuation, we compare the
   quantum states with the classical and
   semiquantal motions in the phase space for some cases.
The semiquantal equation of motion is given by generalized Hamilton-like equations as,
  \begin{eqnarray*}
\frac{d q}{dt} &=& p, \\
\frac{d p}{dt} &=& -q^3+A(t)q-3\hbar \phi q, \\
\frac{d \phi}{dt} &=& 4\phi \pi , \\
\frac{d \pi}{dt} &=& \frac{1}{8\phi^2} -2\pi^2 -\frac{3}{4}q^2 -\frac{1}{4}A(t)-\frac{3}{2}\hbar \phi,  
   \end{eqnarray*}
where the canonical conjugate pair $(\phi,\pi)$ is defined by the quantum 
fluctuation $\Delta q$ and $\Delta p$ as, 
$\Delta q^2=\hbar \phi, \Delta p^2=\hbar(\frac{1}{4\phi} + 4\pi^2\phi)$.
For more details consult \cite{zhang95}.
   It is directly observed that the quantum tunneling phenomenon enhances 
   chaotic motion
   in comparing to the classical and semiquantal trajectories.
   In Fig.8(a) and (b) Poincar\'{e} surface of section of the classical
   trajectories in the phase plane at $M=7$ are shown.
   The stroboscopic plots are taken at $t=2\pi n/\Omega_{1} (n=1,2,...)$,
   due to non time-periodic structure of the Hamiltonian. 
   In the relatively small perturbation strength $\epsilon=0.4$, 
   the trajectories stay the single
   well, and are stable even for the long-time evolution.
   Figure 7(c) and (d) show the Poincar\'{e} section of 
   the semiquantal trajectories for a polychromatically perturbed 
   double-well system with $M=7$ (the stroboscopic plots are taken at 
   $t=2\pi n/\Omega_{1}$ again).
   The semiquantal trajectories for the squeezed quantum coherent state can be obtained by 
   an effective action which includes partial quantum fluctuation to all order in $\hbar$ \cite{pattanayak94}.
   It can be seen that in comparing with ones of classical dynamics 
   the trajectories in the semiquantal dynamics spreads into the opposite well even for the small $\epsilon$.
   This  corresponds to the quantum tunneling phenomenon through the semiquantal dynamics.
   Apparently, the partial quantum fluctuation in the semiquantal
   approximation enhances the the chaotic behavior. 
   Notice that the semiquantal picture breaks down for the irregular quantum states because
   the centroid motion becomes irrelevant.

   In Fig.8(e) and (f) the corresponding coherent state representation for the quantum states
   are shown. It is directly seen that the wave packet spreads over the two-wells and the shape
   is not symmetric.
   Once the wave packet incoherently spreads over the space, it can not return to the initial state
   anymore. 
   We have confirmed that in a case without separatrix (single-well),
   namely the case that $a$ in Eq.~(2) is replaced by $-a$,
   in the classical phase space 
   the coherent oscillations have remained against the relatively large $M$
   and/or $\epsilon$.
   It follows that the full quantum interference suppresses the chaotic behavior 
as seen in  the semiquantual  trajectories.
  
\section{Summary and Discussion}
 We numerically investigated influence of a polychromatic perturbation 
 on wave packet dynamics in one-dimensional double-well potential. 
The calculated physical quantities are the transition rate $P_L(t)$, 
the time-fluctuation $\Delta P_L$,  uncertainty product $\Delta q \Delta p$ 
and phase space portrait. The results we obtained in the present investigation 
are summarized as follows. 

(1) We classified the motions in the parameter space spanned by the
   amplitude and the number of frequency components 
   of the oscillatory driving force, i.e.  {\it coherent motions} and 
{\it irregular motions}.
The critical value $\epsilon_c(M)$ which divides the behavior of $P_L(t)$ 
into regular and irregular motions depends on the number of the frequency
component $M$.  

(2) Within the regular motion range, the period of the tunneling oscillation
is reduced with  increase of the number of colors and/or strength of 
the perturbation. It could be explained by the increase of the instanton tunneling 
rate due to appearance of the stochastic layer near separatrix \cite{instanton}.
In this parameter regime the perturbed instanton picture is one of expression for  
chaos-assisted tunneling \cite{bonci98} and chaos-assisted ionization picture 
reported for some quantum chaos systems \cite{zakrzewski98}.

(3) In the irregular motion in the polychromatically perturbed cases,   
the growth of $\Delta q \Delta p$ initially increases and saturates 
at certain level due to quantum interference. 
On the other hand, in the stochastically perturbed 
case the  uncertainty product grows unboundedly because the external stochasticity
breaks the quantum interference in the internal dynamics.
The growth of $\Delta q \Delta p$ is strongly related to the growth of the
 energy of the wave packet.

(4) It is expected that the quantum fluctuation are always large
for the classically chaotic trajectories compared to the regular ones.
This implies that the quantum corrections to 
the evolution of the phase space fluctuation become more dominant for
classically chaotic trajectories.     

(5) In the semiquantal approximation the partial quantum fluctuation 
 enhances the chaotic behavior, and simultaneously the chaos
 enhances the tunneling and decoherence of the wave packet.
The quantum fluctuation observed in the semiquantal picture 
is suppressed by interference effect in the fully quantum motion.
The semiquantal picture can not apply to the chaos-induced delocalized states.

Furthermore, in the appendices, we gave classical phase space portraits in the 
polychromatically perturbed double-well system and a simple explanation 
for the perturbed instanton tunneling picture for the reduction of the tunneling period
in the coherent motion regime.

\vspace{1cm}
 
Although we  have dealt with quantum dynamics of wave packet 
with paying attention to existence of 
the energetic barrier, we can expect that the similar phenomena
would appear by dynamical barrier in the system. 
The details will be given elsewhere \cite{igarashi06}.

\appendix

\section{Effect of Polychromatic Perturbation on Classical Phase Space Portraits}
We show classical stroboscopic phase space portrait in this appendix 
with paying an attention to the effect of polychromatic perturbation 
on the chaotic behavior. 
In the classical dynamics, such a system shows chaotic behavior by the oscillatory 
force $A(t)$ \cite{igarashi05,igarashi06}.
The Newton's equation of the motion is 
\begin{eqnarray}
  \frac{d^2 q}{dt^2} -A(t)q +q^3=0.
\end{eqnarray}
\noindent
Note that in the monochromatically perturbed case ($M=1$, $A(t)=a-\epsilon \sin \Omega t)$,  
the equation is known as nonlinear Mathieu equation 
which can be derived from surface acoustic wave in piezoelectric solid \cite{konno90} 
and nanomechanical amplifier in micronscale devices \cite{harrington02}.

\subsection{Classical phase space potraits}

In Fig. A.1,  we show the change of the classical stroboscopic phase space portrait 
changing the perturbation parameters. 
Increasing the perturbation strength $\epsilon$ destroys the separatrix and 
forms a chaotic layer in the vicinity of the separatrix.
Needless to say, 
the phenomena have been observed even in the monochromatically perturbed cases \cite{igarashi05}.
In the polychromatically perturbed cases ($M>1$) the smaller the strength $\epsilon$ can generate 
chaotic behavior of the classical trajectories the larger $M$ is \cite{igarashi06}.
It should be emphasized that in the polychromatically perturbed cases 
the width of the chaotic layer grows faster 
than the monochromatically perturbed case as the perturbation strength increases.
As a result, the increase of the color contributes the increase of the width of 
the stochastic layer in the polychromatically perturbed cases. 

   \begin{figure}[!ht]
    \centering
    \includegraphics[clip,scale=0.85]{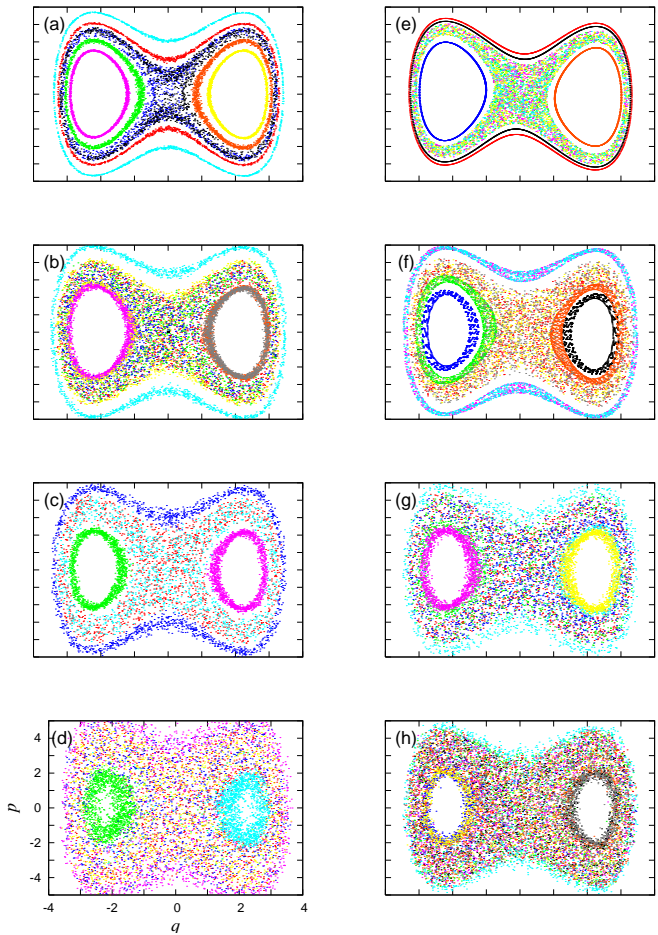}
    \caption{ Stroboscopic classical phase space portraits  
    of the polychromatically perturbed double-well system. 
Cross sections of ($p,q$) space are plotted at 
$2 \pi n/\Omega_1, n=1,2,3,... $.
The $\epsilon-$dependence for $M=5$ is shown 
in (a)$\epsilon=0.1$, (b)$\epsilon=0.3$, (c)$\epsilon=0.4$ and (d)$\epsilon=0.8$.
The $M-$dependence for $\epsilon=0.5$ is shown 
in (e)$M=1$, (f)$M=2$, (g)$M=5$ and (h)$M=8$.
    }
   \end{figure}

Here, 
 we use the increasing rate of infinitesimal displacement 
along the classical trajectory  for the extent of chaotic behavior  
as a finite-time Lyapunov exponent $\lambda_{cl}^{max} $.
We prepare various initial points in the phase space,  
and conveniently adapt a trajectory with maximal increasing rate among 
the ensemble within the finite-time interval as the finite-time Lyapunov
exponent. Note that an exact Lyapunov exponent should be defined for the
long-time limit.
However, the roughly estimated Lyapunov exponent is also useful 
to observe the classical-quantum correspondence.
Figure A.2 shows the $\epsilon-$dependence of classical Lyapunov exponents for various cases 
estimated by the numerical data of the classical trajectories.

   \begin{figure}[!ht]
    \centering
    \includegraphics[clip,scale=1.2]{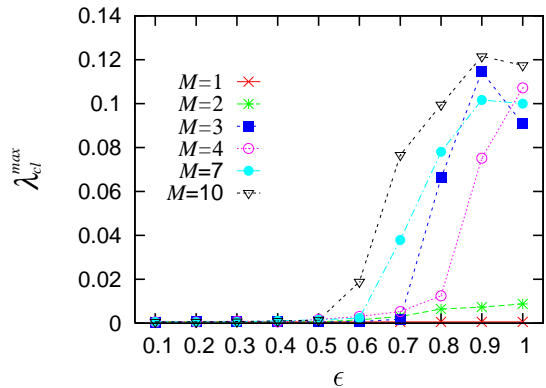}
    \caption{
Maximal Lyapunov exponents $\lambda_{cl}^{max}$ estimated by some classical 
trajectories within the finite-time $t \in [ 0, T]$, where $T$ is tunneling time
given inSubsect.3.1.
    }
   \end{figure}

We can roughly observe a transition from motion of 
KAM system ($\lambda_{cl}^{max} \sim 0$) to chaotic motion ($\lambda_{cl}^{max}  >0$)
as the perturbation strength increases. 
The increasing of the number of color $M$ reduces the value of the critical
perturbation strength $\epsilon_c^{cl}$ of a transition from a motion of KAM system
to fully chaotic motion. 
Roughly speaking, a transition of the classical dynamics corresponds to the transition from coherent motion to 
irregular one in the quantum dynamics. 
We expect that the transition observed in Sect.3 
will corresponds to quantum signatures of 
the KAM transition from the regular to chaotic dynamics.

\subsection{Some considerations to the effect of polychromatic perturbation}

In this subsection, we conceptually consider a role of the polychromatic perturbation different
from monochromatic one.
Note that the change of the number $M$ of colors also changes the qualitative nature of 
the underlying dynamics because $M$ corresponds to the effective 
number of DOF under some conditions \cite{ikeda93,yamada99}. 
In the our model, when the number of DOF of the total system is more than four, 
i.e. $M \geq 3$, the classical trajectories can diffuse along 
the stochastic layers of many resonances that cover the whole phase space 
if the trajectory starts in the vicinity of a nonlinear resonance. 
The number of resonances increases rapidly with DOF, 
changing the characteristic of population transfer from bounded to diffusive.
Such a global instability is known as Arnold diffusion in nonlinear Hamiltonian
system with many DOF \cite{chirikov79}. 
The effect of Arnold diffusion in quantum system is not trivial and the study
just has started recently \cite{demikhovskii02}. 
The more detail is out of scope of this paper.

Moreover, we can regard the time-dependent model of Eq.(1) as nonautonomous approximation
for an autonomous model, consisting of the double-well 
system coupled finite number $M$ of harmonic oscillators with the incommensurate 
frequencies $\{ \Omega_j, j=1,1,...,M \}$.    
 It is worth noting that the linear oscillators can be identified with a highly
excited quantum harmonic oscillators, which all phonon modes are excited around
Fock states with large quantum numbers.  Then the above model can be regarded 
as a double-well system coupled with $M$ phonon modes. Without the interaction 
with the phonon modes the Gaussian wavepacket remains the coherent motion.
 Then the number of DOF of total system is $1+M$ and 
the number of the frequency components $M$ corresponds to the that of 
the highly excited quantum harmonic oscillators.  
The detail of the correspondence is given in Ref.\cite{yamada99}.

In quantum chaotic system with finite and many DOF we expect occurrence of 
a dissipative behavior. 
For example, 
we consider simulated light absorption by coupling a system in the ground
state with radiation field. Then stationary one-way energy transport from 
photon source to the system can be interpreted as occurrence of quantum 
irreversibility in the total system. Such a irreversibility  is called 
 chaos-induced dissipation in quantum system with more than two DOF \cite{ikeda93}.
In this sense, we can expect occurrence of the one-way transport phenomenon
 in the delocalized state in the irregular motion phase if it couples with 
the other DOF in the ground state as seen in Ref.\cite{yamada99}.

\section{Perturbed Instanton Tunneling}
In this appendix, we consider the reduction of the tunneling period as the 
perturbation strength increases in the coherent oscilation regime $\epsilon <\epsilon_c$, 
based on a perturbed instanton tunneling. 
In a double-well system with dipole-type interaction, $q\sin(\Omega t)$,  
the energitical barrier tunneling between the symmetric double-well 
can be explained by a three-state model or  chaos-assisted tunneling (CAT)
\cite{tomsovic94,kohler98,bonci98,zakrzewski98}.  
The three states that 
take part in the tunneling are a doublet of quasi-degenerate states with
opposite parity, localized in the each well, and a third state localized 
in the chaotic layer around the separatrix. 

However, note that less attention has been paid to tunneling in 
KAM system while chaotic dynamics has been modeled by multi-level Hamiltonian
and random matrix model to describe the chaos-assisted tunneling.
We give an expresion of the tunneling amplitude in  "chaos-assisted instanton 
tunneling" firstly  proposed by Kuvshinov {\it et al} for a Hamiltonian system with time-periodic
perturbation \cite{instanton}.
Let us consider only  monochromatically perturbed case ($M=1$ in Eq.(2), $\Omega=\Omega_1$)
because the separatrix destruction mechanism by the time-periodic perturbation has an universality
although our system is different from their one. 
. 

\subsection{A discrete mapping}
Indeed, trajectories in the neighborhood of the separatrix of the system
are well reproduced by the whisker map of the system.
Whisker map is a map of the energy change $I_n$ and phase change $\phi_n$ of a trajectory 
in the neighborhood of the separatrix for each of its motion during one period of the perturbation, 
i.e. action-angle variable.
Moreover, if we linearize the whisker map which describes the behaviors of the trajectories 
in the neighborhood of  the fixed point, we can 
obtain the following standard map, 
\begin{eqnarray}
\left\{ 
\begin{array}{ll}
I_{n+1} & =   I_{n} -K(\epsilon,\Omega) \sin \phi_n  \\
\phi_{n+1} & =  \phi_n + I_{n+1}  \\
\end{array} \right.
\end{eqnarray}
\noindent
, where $K(\epsilon,\Omega)$ is a nonlinear parameter of local instability that 
the exact function form which is not essential for our purpose. 
$K$ increases with $\epsilon$, and $K\geq 1$ means that the dynamics of the system is locally unstable.
A comparison has done between the whisker map and the strobe plots in the 
time continuous version by Yamaguchi \cite{yamaguchi85}.
The form of the mapping is convenient for the estimate of the width of the
stochastic layer.

Kuvshivov {\it et al} has given the estimated parameter of 
local instability, width of stochastic layer, and correlator for 
perturbed instanton solution for the perturbed pendulum system \cite{instanton}.

\subsection{Action of the instanton tunneling}
The perturbation destroys separatrix of the unperturbed system 
and the stochastic layer appears. 
In the regular motion denoted by the circles in Fig.3, 
classical chaos can increase the rate of instanton tunneling due to 
appearance of the stochastic layer near separatrix of the unperturbed system.
As a result the frequency of time-dependence $P_L(t)$ increases as
the classical chaos becomes remarkable in the parameter regime.
Note that the perturbed instanton tunneling picture disappears in the strongly perturbed regime 
due to the delocalization of wavepacket.

Here we give only relation between the width of the stochastic layer and the tunneling 
amplitude in terms of path integral in imaginary time $\tau$,  found by Kuvshivov {\it et al}.
Tunneling amplitude between the two wells 
in the perturbed system can be given by integration over energy of 
tunneling amplitude $A_{tun}$ in unperturbed system as, 
\begin{eqnarray}
A_{tun} & =& \int_0^{\Delta H} dE \int_{q(\tau)=-q_0}^{q(\tau)=q_0} D[q(\tau,E)] \exp\{ -S[q(\tau, E)] \}, 
\end{eqnarray}
\noindent
where $S[q(\tau, E)] $ denotes the Euclidian action.
 $\Delta H \equiv 2|H_s-H_b|$ denotes the width of stochastic layer, where 
$H_s$  and $H_b$ are the energy of the unperturbed system on the separatrix and 
on the bound of stochastic layer, respectively.
 $q(\tau, E)$ is classical solution of Euclidian equation of motion. 
The contribution of the chaotic instanton solution are taken into account 
by means of integration over $E$ which is energy of the instanton.

The perturbed instanton solutions correspond to the motions in vicinity of 
the separatrix inside the layer. 
The only manifestation of the perturbation in this approximation is the appearance 
of a number of  additional solutions of the Euclidian equation of motion with energy
close to the energy of the unperturbed one-instanton solution 
inside the stochastic layer. 
Accordingly, we can expect that the appearance of the stochastic layer enhances
the tunneling rate as reported in the other systems \cite{instanton}.
However, we have to have in mind that the result is obtained in the first order on coupling 
constant $\epsilon$ of the time-periodic perturbation and does not take into
account the structure of stochastic layer. 
The approximation is valid if the layer is narrow 
by neglecting the higher order resonances in the phase space.
For the more details of the perturbed instanton see Ref.\cite{instanton}.
The increasing of the tunneling amplitude is directly related to energy splitting
$\Delta E_{01}$ between  the quasi-degenerate ground Floquet states.

As seen in Fig.A.1, the increasing of number of color $M$ can enhance 
the width of the stochastic layer with the perturbation strength being kept at a 
constant value. 
The theoretical explanation for the reduction of the tunneling period with the number of color 
is open for further study.

We expect that in the double-well system under the polychromatic perturbation
this numerical study will be useful for the analytical derivation 
of "reduction of tunneling period" and "critical strength of a transition from
localized to delocalized behavior of wavepacket" by extension of the 
monochromatically perturbed case. 
The chaos assisted instanton theory might be applicable if we will exactly 
estimate the width of the stochastic layer in the system under the polychromatic
perturbation.





\begin{thebibliography}{00}
\bibitem{duffing} 
We have confirmed that result in the polychromatically perturbed Duffing-like model 
was qualitatively same as the present one.
\end{thebibliography}
\end{document}